\documentstyle[prd,aps]{revtex}
\input epsf

\def\Journal#1#2#3#4{{#1} {\bf #2}, #3 (#4)}
\def\MPL{Mod. Phys. Lett. A}

\def\NPB{Nucl. Phys. B}
\def\NPBOLD{Nucl. Phys.}

\def\PLB{{Phys. Lett.} B}

\def\PLBOLD{Phys. Lett.}
\def\PRL{Phys. Rev. Lett.}
\def\RMP{Rev. Mod. Phys.}
\def\PRC{Phys. Rev. C}
\def\PRD{Phys. Rev. D}

\def\PTP{Prog. Theor. Phys.}
\def\JHEP{JHEP}
\def\APJ{Astrophys. J.}

\def\JETPUSSR{JETP (USSR)}
\def\ZETP{Zh. Eksp. Teor. Piz.}

\def\mapgeq{\mathbin{\lower.3ex\hbox{$\buildrel>\over{\smash{\scriptstyle\sim}\vphantom{_x}}$}}}
\def\mapleq{\mathbin{\lower.3ex\hbox{$\buildrel<\over{\smash{\scriptstyle\sim}\vphantom{_x}}$}}}
\def\mapgeqeq{\mathbi{\lower.3ex\hbox{$\buildrel>\over{\smash{\scriptstyle\approx}\vphantom{_2}}$}}}
\def\mapleqeq{\mathbin{\lower.3ex\hbox{$\buildrel<\over{\smash{\scriptstyle\approx}\vphantom{_2}}$}}}

 \mathchardef\#="0023
 \mathchardef\$="0024
 \mathchardef\%="0025
 \mathchardef\ddash="705C
 
 \mathchardef\lwavy="336E
 \mathchardef\rwavy="336F
 \mathchardef\biglwavy="331A
 \mathchardef\bigrwavy="331B
 \mathchardef\bigglwavy="3328
 \mathchardef\biggrwavy="3329
 \mathchardef\littlesum="0350

\tighten
\draft
\begin{document} 
\bibliographystyle{prsty}

\title{
Comment on Neutrino Masses and Oscillations \\
in an $SU(3)_L \times U(1)_N$ Model \\
with Radiative Mechanism}

\author{
Teruyuki Kitabayashi
\footnote{E-mail:teruyuki@post.kek.jp}
}
\address{\vspace{4mm}
{\sl Accelerator Engineering Center,}\\ 
{\sl Mitsubishi Electric System \& Service Engineering Co.Ltd.,} \\
{\sl 2-8-8 Umezono, Tsukuba, Ibaraki 305-0045, Japan.}
}
\date{\small March, 2001}
\maketitle

\begin{abstract}
We discussed how neutrino masses and oscillations are radiatively generated in an $SU(3)_L$ $\times$ $U(1)_N$ gauge model with a symmetry based on $L_e-L_\mu-L_\tau$ ($\equiv$$L^\prime$). The model is characterized by lepton triplets $\psi^i=(\nu^i,\ell^{-i},E^{-i})$, where $E^{-i}$ are negatively charged heavy leptons, an $SU(3)_L$ triplet Higgs scalar $\xi$ and a singlet Higgs scalar $k^{++}$. These Higgs scalars can be interpreted as a Zee's and Zee-Babu's scalar for radiative mechanisms. We demonstrated that the mass hierarchy of $\Delta m_{atm}^2$ $\gg$ $\Delta m_\odot^2$ arise as a consequence of the dynamical hierarchy between $L^\prime$-conserving one-loop effects and $L^\prime$-violating two-loop effects, and our model is relevant to yield quasivacuum solution for solar neutrino problem. 
\end{abstract}
\pacs{PACS: 12.60.-i, 13.15.+g, 14.60.Pq, 14.60.St\\
Keywords: neutrino mass, neutrino oscillation, radiative mechanism, lepton triplet}

\vspace{2mm}
There is definitive evidence for neutrino oscillations from atmospheric and solar neutrino observations. For the atmospheric neutrino oscillations, the recent SuperKamiokande (SK) data indicates that the observed deficit of $\nu_\mu$ is due to the $\nu_\mu$ $\leftrightarrow$ $\nu_\tau$ oscillation \cite{Kamiokande,RecentSK}, while for the solar neutrino oscillations \cite{Solar}, the SK, Homestake \cite{Homestake}, SAGE \cite{SAGE}, GALLEX \cite{GALLEX} and GNO \cite{GNO} data indicate the $\nu_e$ $\leftrightarrow$ $\nu_\mu,\nu_\tau$ oscillation. The existence of these neutrino oscillations implies the neutrinos are massive particles \cite{MassiveNeutrino}. The mass squared differences for atmospheric oscillations $\Delta m_{atm}^2$ is measured as $\Delta m_{atm}^2$ $\sim$ $3\times10^{-3}$ eV$^2$ \cite{RecentK2K}. On the other hand, there are some solutions to explain the observed solar neutrino oscillation data as (1) $\Delta m_\odot^2$ $\sim$ $10^{-5}$ eV$^2$ for the LMA solution, (2) $\Delta m_\odot^2$ $\sim$ $10^{-6}$ eV$^2$ for the SMA solution, (3) $\Delta m_\odot^2$ $\sim$ $10^{-7}$ eV$^2$ for the LOW solution, (4) $\Delta m_\odot^2$ $\sim$ $10^{-10}$ eV$^2$ for the VO solution and recently proposed (5) $\Delta m_\odot^2$ $\sim$ $10^{-9}$ eV$^2$ for the quasi-VO (QVO) solution \cite{QVO}. To sum up, we have $\Delta m_{atm}^2$ $\sim$ $10^{-3}$ eV$^2$ and $\Delta m_\odot^2$ $\mapleq$ $10^{-5}$ eV$^2$ indicating the hierarchy of $\Delta m_{atm}^2$ $\gg$ $\Delta m_\odot^2$ exist. In the theoretical view, this mass hierarchy suggests that the neutrino mass matrix has bimaximal structure \cite{NearlyBiMaximal,Mixing}.

Recently, radiative mechanisms to generate tiny neutrino masses and oscillations in $SU(3)_L$ $\times$ $U(1)_N$ gauge models \cite{SU3U1,SU3U1Topics,Lately} with the $L^\prime$ symmetry have been extensively studied \cite{Kita00a,Kita00b,Kita01a}. Here $L^\prime$ $\equiv$ $L_e-L_\mu-L_\tau$ is a new lepton number and the conservation of this quantum number is one of the possibilities of the origin of the bimaximal structure \cite{Lprime,EarlierLprime}. Three $SU(3)_L$ $\times$ $U(1)_N$ gauge models are used to accomodate such radiative mechanisms. Each of the $SU(3)_L$ $\times$ $U(1)_N$ models can be distinguished by the lepton triplets $\psi^i$ ($i=1,2,3$) in the models: (a) $\psi^i$ = $(\nu^i,\ell^i,\omega^{0i})$ model \cite{Kita00a}, (b) $\psi^i$ = $(\nu^i,\ell^i,\kappa^{+i})$ model \footnote{The model with $\psi^i$ = $(\nu^i.\ell^i,\ell^{+i})$ has been lately examined to yield tiny neutrino masses and observed neutrino oscillations \cite{Lately}.} \cite{Kita00b} and (c) $\psi^i$ = $(\nu^i,\ell^i,E^{-i})$ model \cite{Kita01a}, where $\omega^{0i}$, $\kappa^{+i}$ and $E^{-i}$ are denoted by electrically neutral heavy leptons, positively charged heavy leptons and negatively charged heavy leptons, respectively. In the model (a) and the model (b), the atmospheric neutrino oscillations are generated by a one-loop radiative mechanism with $L^\prime$-conserving interactions \cite{1-loopLprime}, and the solar neutrino oscillations are induced from a two-loop rediative mechanism with $L^\prime$-violating interactions \cite{2-loopLprime}. Consequently, the mass hierarchy of $\Delta m_{atm}^2$ $\gg$ $\Delta m_\odot^2$ is explained as a result of the smallness of the two-loop effects compared with one-loop effects \cite{1loop2loop,1loop2loopNew}. On the other hand, in the model (c), there is no one-loop interaction and both of the atmospheric and solar neutrino oscillations come from two-loop radiative effects \cite{2-loopLprime}. The mass hierarchy of $\Delta m_{atm}^2$ $\gg$ $\Delta m_\odot^2$ is related to the dynamical hierarchy of the $L^\prime$-conserving and $L^\prime$-violating two-loop interaction effects.
 
 In this article, we show that it is possible to construct the other $SU(3)_L \times U(1)_N$ model with lepton triplets $\psi^i$ = $(\nu^i,\ell^i,E^{-i})$ \cite{HeavyE}. The model has similar particle content to the model (c); however, one-loop interactions also exist and neutrino masses are induced by the $L^\prime$-conserving one-loop radiative mechanism as well as the $L^\prime$-violating two-loop radiative mechanism.
 
The particle content in our $SU(3)_L \times U(1)_N$ gauge model is summarized as follows:

\begin{eqnarray}
\psi^{i=1,2,3}_L=\left(\nu^i,\ell^i,E^i\right)_L^T:\left(\textbf{3},-2/3\right),\quad
\ell^{1,2,3}_R                                    :\left(\textbf{1},-1 \right), \quad  
E^{1,2,3}_R                                       :\left(\textbf{1},-1 \right),
\label{Eq:Leptons}
\end{eqnarray}
in the lepton sector, where we have denoted $E^{-i}$ by $E^i$,
\begin{eqnarray}
&&Q^{1}_L=\left(u^1,d^1,d^{\prime 1} \right)_L^T    :\left(\textbf{3},0 \right),\quad
  Q^{i=2,3}_L=\left(d^i,-u^i,u^{\prime i}\right)_L^T:\left(\textbf{3}^\ast,1/3\right),
\nonumber \\
&&u^{1,2,3}_R     :\left( \textbf{1}, 2/3 \right),\quad 
  d^{1,2,3}_R     :\left( \textbf{1},-1/3 \right),\quad 
  u^{\prime 2,3}_R:\left( \textbf{1}, 2/3 \right),\quad
  d^{\prime 1  }_R:\left( \textbf{1},-1/3 \right),
\label{Eq:Quarks}
\end{eqnarray}
in the quark sector, and
\begin{eqnarray}
&&\eta=\left(\eta^0,\eta^-,\overline{\eta}^-\right)^T:(\textbf{3},-2/3), \quad
  \rho=\left(\rho^+,\rho^0,\overline{\rho}^0\right)^T:(\textbf{3}, 1/3),
\nonumber \\
&&\chi=\left(\chi^+,\overline{\chi}^0,\chi^0 \right)^T:(\textbf{3},1/3), \quad  
  \xi =\left(\xi^{++},\overline{\xi}^+,\xi^+ \right)^T:(\textbf{3},4/3),
\nonumber \\
&&k^{++}:(\textbf{1},2),
\label{Eq:Higgs}
\end{eqnarray}
in the Higgs sector, where the quantum numbers are specified in the parentheses by $(SU(3)_L,U(1)_N)$. Let $N/2$ be the $U(1)_N$ quantum number, then the hypercharge ($Y$) and electric charge ($Q_{e}$) are given by $Y=\lambda^8/\sqrt{3}+N$ and $Q_e=(\lambda^3+Y)/2$, respectively, where $\lambda^a$ is the $SU(3)$ generator with Tr($\lambda^a \lambda^b$) = $2\delta^{ab}$ ($a,b$ = $1,...,8$). Three Higgs triplets $\eta, \rho$ and $\chi$ are the minimal set to generate masses of quarks and leptons in $SU(3)$ $\times$ $U(1)_N$ models. An additional Higgs triplet $\xi$ is introduced as a triplet version of the Zee scalar to realize the one-loop radiative mechanism \cite{1-loop} and an additional Higgs singlet $k^{++}$ is introduced to realize the two-loop radiative mechanism \cite{2-loop}.

Here, we introduce two constraints to obtain the relevant Yukawa interactions. The first is the $L^\prime$ $\equiv$ $L_e-L_\mu-L_\tau$ conservation imposed on our interactions to reproduce the observed atmospheric neutrino oscillations as mentioned. The $L^\prime$ assignment is shown in Table \ref{Tab:Lnumber}. The second is the discrete symmetry based on $Z_4$ to suppress unwanted flavor-changing-neutral-currents (FCNC) interactions in the quark sector and the lepton sector. In the quark sector, there are quarks with the same charge, thus, quark mass terms can be generated by $\rho$ and $\chi$ between $Q_L^1$ and down-type quarks and by $\rho^\dagger$ and $\chi^\dagger$ between $Q_L^{2,3}$ and up-type quarks. FCNC is induced from these interactions \cite{FCNCSU3}. Also, in the lepton sector, $\ell^i$ and $E^i$ ($i=1,2,3$) has the same chage and the similar FCNC problem can occur. To avoid such interactions, Yukawa interactions must be constrained such that a quark (lepton) flavor gains a mass from only one Higgs scalar \cite{FCNC}. These situations can be realized by introducing the following $Z_4$ symmetry into the model:
$\psi_L^{1,2,3} \rightarrow i\psi_L^{1,2,3}$,
$\ell_R^{1,2,3} \rightarrow \ell_R^{1,2,3}$,
$E_R^{1,2,3}    \rightarrow -iE_R^{1,2,3}$,
$Q_L^1          \rightarrow  iQ_L^1$,
$Q_L^{2,3}      \rightarrow -iQ_L^{2,3}$,
$u_R^{1,2,3}      \rightarrow   u_R^{1,2,3}$,
$d_R^{1,2,3}      \rightarrow   d_R^{1,2,3}$,
$u_R^{\prime 2,3} \rightarrow  iu_R^{\prime 2,3}$,
$d_R^{\prime 1}   \rightarrow -id_R^{\prime 1}$,
$\eta             \rightarrow  i\eta$,
$\rho             \rightarrow  i\rho$,
$\chi             \rightarrow  -\chi$,
$\xi              \rightarrow  -\xi$,
and 
$k^{++}           \rightarrow  ik^{++}$.
 
With these constraints, the Yukawa interactions are given by
\begin{eqnarray}
-{\mathcal{L}}_Y &=& 
     \epsilon^{\alpha\beta\gamma}\sum_{i=2,3}f_{[1i]}
     \overline{\left(\psi_{\alpha L}^1 \right)^c} \psi_{\beta L}^i \xi_\gamma
   + \sum_{i=1,2,3} \overline{\psi_L^i}
     \left( f_\ell^i \rho \ell_R^i + f_E^i \chi E_R^i \right)
\nonumber \\
&& + \sum_{i,j=2,3}f_k^{ij}\overline{(\ell_R^i)^c} E_R^j k^{++} 
   + \overline{Q_L^1}\left( \eta U_R^1 + \rho D_R^1 + \chi D_R^{\prime 1} \right)
\nonumber \\
&& + \sum_{i=2,3} \overline{Q_L^i}
   \left( \eta^\ast D_R^i + \rho^\ast U_R^i + \chi^\ast U_R^{\prime i} \right)
   + (h.c.),
\label{Eq:Yukawa}
\end{eqnarray}
where $f$'s are Yukawa couplings with the relation $f_{[ij]}=-f_{[ji]}$ demanded by the Fermi statics, and right-handed quarks are denoted by $U_R^i=\sum_{j=1}^3 f_{uj}^i u_R^j$, $D_R^i=\sum_{j=1}^3 f_{dj}^i d_R^j$, $U_R^{\prime i}=f_{u^\prime 2,3}^i u_R^{\prime 2,3}$ and $D_R^{\prime 1}=f_{d^\prime 1}^1 d_R^{\prime 1}$. For simplicity, we have assumed diagonal mass terms for the leptons. Note that there is no term which can induce FCNC interactions such as $\overline{Q_L^1} \chi D_R^1$, $\overline{Q_L^1} \rho D_R^{\prime 1}$, $\overline{Q_L^{2,3}} \chi^\ast U_R^{2,3}$, $\overline{Q_L^{2,3}} \rho^\ast U_R^{\prime 2,3}$, $\overline{\psi_L^i} \chi \ell_R^i$ and $\overline{\psi_L^i} \rho E_R^i$. 

The Higgs interactions are given by self-Hermitian terms of $\phi_\alpha\phi^\dagger_\beta$ ($\phi$ = $\rho$, $\eta$, $\chi$, $\xi$, $k^{++}$), and two types of non-self-Hermitian Higgs potentials:
\begin{eqnarray}
V_0 &=& \lambda_0 \epsilon^{\alpha\beta\gamma} \eta_\alpha \rho_\beta \chi_\gamma
	 + \lambda_1 (\eta^\dagger \rho)(\xi^\dagger \chi)
	 + \lambda_2 (\eta^\dagger \chi)(\xi^\dagger \rho) + (h.c.),
\nonumber \\
V_b &=& \mu_b \xi^\dagger \eta k^{++} + (h.c.),
\label{Eq:HiggsPotentials}
\end{eqnarray}
where $\lambda_{0,1,2}$ stands for $L^\prime$-conserving coupling constants and $\mu_b$ denotes the $L^\prime$-violating mass scale. The interaction of the $\eta\rho\chi$ type in Eq.(\ref{Eq:HiggsPotentials}) is a guarantee of the orthogonal choice of vacuum expectation values (VEVs) for three Higgs scalars, $\eta$, $\rho$ and $\chi$ as $\langle 0 \vert \eta\vert  0 \rangle=(v_\eta,0,0)^T$, $\langle 0 \vert \rho\vert  0 \rangle=(0,v_\rho,0)^T$, and $\langle 0 \vert \chi\vert  0 \rangle=(0,0,v_\chi)^T$, respectively.
 
  We note that there are two main differences between the model (c) discussed in Ref.\cite{Kita01a} and the model in this article (current model). The first is the absence of an $SU(3)_L$ singlet Higgs scalar $k^{\prime ++}$ in the current model. The model (c) has two Zee-Babu type Higgs scalars called $k^{++}$ and $k^{\prime ++}$, which are needed to realize $L^\prime$-conserving and $L^\prime$-violating two-loop interactions. However, in the current model, only one Higgs $k^{++}$ is introduced and no additional singlet Higgs is needed because $L^\prime$-conserving one-loop effects will serve as the $L^\prime$-violating two-loop effects in the model (c). The second is the different implementation of the discrete symmetry into the models. The discrete symmetry based on $Z_2$ is required in model (c) to avoid the FCNC interactions and to prevent the realization of the one-loop effects. Meanwhile, in the current model, the discrete symmetry based on $Z_4$ is introduced and the one-loop effects are allowed.

Now, let us demonstrate how radiative corrections induce neutrino masses in our model. The Yukawa interaction denoted by ${\mathcal{L}}_Y$ and $L^\prime$-conserving Higgs potential $V_0$ work together to generate one-loop interactions as shown in Fig. \ref{Fig:oneloop}, also ${\mathcal{L}}_Y$ and $L^\prime$-violating Higgs potential $V_b$ yield two-loop interactions as shown in Fig. \ref{Fig:twoloop}. From the one-loop diagrams, we obtained the following Majorana neutrino masses:
\begin{eqnarray}
m_{1i}^{(1)} &=& f_{[1i]}
    \Biggl[ 
	\lambda_1 
    \frac{m_{\ell^i}^2 F\left(m_{\ell^i}^2,m_{\xi^+}^2,m_{\rho^+}^2 \right)
        - m_e^2        F\left(m_e^2,       m_{\xi^+}^2,m_{\rho^+}^2 \right)
    }{v_\rho^2}
\nonumber \\
&& +\lambda_2
    \frac{m_{E^i}^2 F\left(m_{E^i}^2,m_{\overline{\xi}^+}^2,m_{\chi^+}^2 \right)
        - m_{E^1}^2 F\left(m_{E^1}^2,m_{\overline{\xi}^+}^2,m_{\chi^+}^2 \right)
    }{v_\chi^2}
    \Biggl] v_\eta v_\rho v_\chi,
\label{Eq:MajoranaMass1}
\end{eqnarray}
where
\begin{eqnarray}
F(x,y,z)&=&\frac{1}{16\pi^2}
    \Biggl[\frac{x\ln x}{(x-y)(x-z)}
	      +\frac{y\ln y}{(y-x)(y-z)}
		  +\frac{z\ln z}{(z-y)(z-x)}
    \Biggl],
\label{Eq:Fxyz}
\end{eqnarray}
and, from the two-loop diagrams, we obtain:
\begin{eqnarray}
m_{11}^{(2)} &=& -2\sum_{i,j=2,3}\lambda_2 f_{[1i]}f_{[1j]}f_k^{ij}
                  \mu_b m_{\ell^i}m_{E^j} v_\rho v_\chi I^{(2)} 
\label{Eq:MajoranaMass2}
\end{eqnarray}
with
\begin{eqnarray}
I^{(2)}
    &=&\frac{G(m_{\ell^i}^2,m_{\xi^+}^2) \left[G(m_{E^j}^2,m_{\overline{\eta}^-}^2)
			                                  -G(m_{E^j}^2,m_{\overline{\xi}^+}^2)
		                               \right]
			}{m_k^2 \left(m_{\overline{\eta}^-}^2-m_{\overline{\xi}^+}^2\right)},
\nonumber \\
G(m_a^2,m_b^2)
    &=&\frac{1}{16\pi^2}
	   \frac{m_a^2\ln(m_a^2/m_k^2)-m_b^2\ln(m_b^2/m_k^2)}{m_a^2-m_b^2},
\label{Eq:I2}
\end{eqnarray}
where the relation of $m_k \gg $ (masses of other particle) has been used. The outline of the derivation of the two-loop integral, Eq.(\ref{Eq:I2}), is shown in the Appendix of Ref.\cite{Kita00a}. 

The neutrino mass matrix is composed of these Majorana masses as 
\begin{equation}
M_\nu=\left(
\begin{array}{ccc}
m_{11}^{(2)}&m_{12}^{(1)}&m_{13}^{(1)}\\
m_{12}^{(1)}&0&0\\
m_{13}^{(1)}&0&0\\
\end{array}
\right),
\label{Eq:MassMatrix}
\end{equation}
from which we find the following relations for the neutrino oscillations in our $SU(3)_L$ $\times$ $U(1)_N$ model
\begin{eqnarray}
\Delta m_{atm}^2 = m_{12}^{(1)2}+m_{13}^{(1)2} (\equiv m_\nu^2), \quad
\Delta m_\odot^2 = 2m_\nu \vert m_{11}^{(2)}\vert .
\label{Eq:DeltaM}
\end{eqnarray}
The bimaxmal structure of $M_\nu$ is realized by requiring that $\vert m_{12}^{(1)}\vert\sim \vert m_{13}^{(1)}\vert$, thereby, leading to $m_{E^2}\sim m_{E^3}$ or $m_{E^2,E^3}\ll m_{E^1}$ because the charged lepton contributions are to be neglected in our case.  We assume that $m_{E^2}\sim m_{E^3}$ in our analysis.

In order to see that our result, Eq.(\ref{Eq:DeltaM}), really regenerates the observed neutrino oscillations, we make the following assumptions on relevant free parameters in the same way as those in Ref.\cite{Kita00b}:
(1) $v_\eta=v_w/20$, $v_\rho=v_w$ and $v_\chi=10v_w$, where $v_w=(2\sqrt{2}G_F)^{-1/2}=$174 GeV,
(2) $m_{\xi} \sim m_\rho = v_w$, $m_{\chi,k} = 10v_w$, $m_{E^2,E^3} = ev_\chi$ to enhance the bimaximal mixing and $m_{E^1} = 0.9m_{E^2,E^3}$ to contribute to $\Delta m_{atm}^2$, where $e$ stands for the electromagnetic coupling,
(3) $f_{[1i]} \sim 10^{-7}$, $\lambda_1=\lambda_2=f_k^{ij}=1$ and $\mu_b=ev_\chi$,
where $f_{[1i]}$ is determined by $\Delta m_{atm}^2$ = $m_{12}^{(1)2}+m_{13}^{(1)2}$ = $3.0\times 10^{-3}$ eV$^2$. From numerical calculations of Eq.(\ref{Eq:DeltaM}), we find $f_{[1i]}=0.93\times 10^{-7}$ eV$^2$, which reproduce $\Delta m_{atm}^2 = 3.0\times 10^{-3}$  eV$^2$ and $\Delta m_\odot^2 = 0.91\times 10^{-9}$ eV$^2$. As a result, the mixing angle $\vartheta$ for atmospheric neutrinos defined by $\cos\vartheta=m_{12}^{(1)}/m_\nu$ is computed to yield $\sin^22\vartheta = 0.93$, where the charged lepton contributions to $\Delta m^2_{atm}$ give the deviation form $\sin^22\vartheta=1$ for the bimaximal mixing case. The estimated $\Delta m_\odot^2$ lies in the allowed region of the QVO solution to the solar neutrino problem.

Summarizing our discussion, we have constructed an $SU(3)_L$ $\times$ $U(1)_N$ gauge model with lepton triplets $\psi^i=(\nu^i,\ell^{-i},E^{-i})$, where $E^{-i}$ are negatively charged heavy leptons. This model has a triplet version of the Zee scalar $\xi$ and a singlet as the Zee-Babu scalar $k^{++}$. Owing to the existence of these scalars, our $SU(3)_L \times U(1)_N$ model is capable of generating tiny neutrino masses by the radiative mechanism. The atmospheric neutrino oscillation is related to $L^\prime$-conserving one-loop interactions, while the solar neutrino oscillation is related to $L^\prime$-violating two-loop interactions, where $L^\prime \equiv L_e-L_\mu-L_\tau$. As a result, the bimaximal structure of the neutrino mass matrix is enhanced by the approximate degeneracy between masses of heavy leptons of $E^2$ and $E^3$.  The observed mass hierarchy of $\Delta m_{atm}^2$ $\gg$ $\Delta m_\odot^2$ is explained by the difference between one-loop and two-loop effects. From our numerical estimate, our model reproduces the observed neutrino oscillation data $\Delta m_{atm}^2 = 3.0\times 10^{-3}$ eV$^2$ with $\sin^2\vartheta = 0.93$ and $\Delta m_\odot^2 = 0.91\times 10^{-9}$ eV$^2$. Our model is, thus,  relevant to yield quasivacuum solution for solar neutrino problem. 

\bigskip
\noindent

The author would like to thank Prof. M.Yasu\`{e} for many helpful suggestions, useful comments and a careful reading of this article.


\noindent
\begin{center}
\textbf{Table Captions}
\end{center}
\begin{table}[ht]
    \caption{\label{Tab:Lnumber}$L$ and $L^\prime$ quantum number.}
\end{table}

\bigskip
\noindent

\begin{center}
\textbf{Figure Captions}
\end{center}
\begin{figure}[ht]
  \caption{$L^\prime$-conserving one-loop diagrams.}
\label{Fig:oneloop}
\end{figure}
\begin{figure}[ht]
  \caption{$L^\prime$-violating two-loop diagrams.}
\label{Fig:twoloop}
\end{figure}

\bigskip
\bigskip
\bigskip
\centerline{TABLE \ref{Tab:Lnumber}. $L$ and $L^\prime$ quantum number.}
\begin{table}[ht]
    \begin{center}
    \begin{tabular}{ccccc}
    \hline
        Fields     & $\eta,\rho,\chi,\xi$ & $\psi_L^1,\ell_R^1,E_R^1$
                   & $\psi_L^{2,3},\ell_R^{2,3},E_R^{2,3}$ & $k^{++}$ \\
    \hline
        $L$        & 0 & 1 &  1 & -2 \\
    \hline
        $L^\prime$ & 0 & 1 & -1 & -2 \\
    \hline
    \end{tabular}
    \end{center}
\end{table}
\newpage

\centerline{\epsfbox{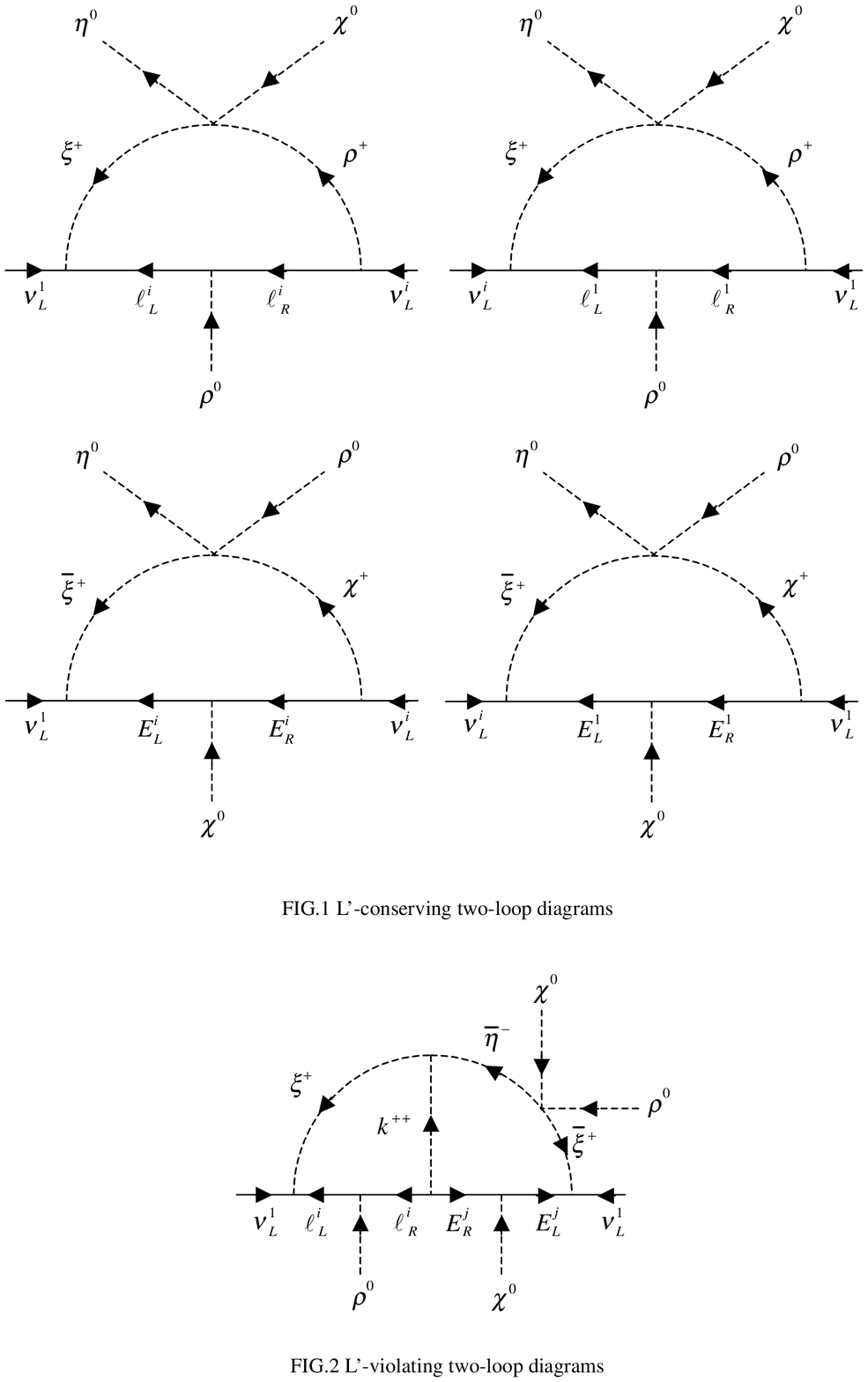}}


\begin{references}
\bibitem{Kamiokande}      
    Super-Kamiokande Collaboration, Y. Fukuda. et. al., \Journal{\PRL}{81}{1562}{1998};
    \Journal{\PLB}{433}{9}{1998}; \Journal{\PLB}{436}{33}{1998};
    N. Fornengo, M.C. Gonzalez-Garcia and J.W.F. Valle, \Journal{\NPB}{580}{58}{2000};
    T. Kajita and Y. Totsuka, \Journal{\RMP}{73}{85}{2001}.
\bibitem{RecentSK}
    Y. Takeuchi, Talk given at 
    \textit{the 30th Int. Conf. on High Energy Physics (ICHEP2000)}, 
    27 Jul. - 2 Aug., Osaka, Japan (http://ichep2000.hep.sci.osaka-u.ac.jp/scan/0728/pa08/takeuchi/index.html);
    M. Smy, Talk given at 
    \textit{the 2nd Workshop on Neutrino Oscillations and Their Origin (NOON 2000)}, 
    6 - 8 Dec., Univ. of Tokyo, Tokyo, Japan (http://www-sk.icrr.u-tokyo.ac.jp/noon/2/transparency/index.html).
\bibitem{Solar}
    J.N. Bahcall, P.I. Krastev and A.Yu.Smirnov, 
    \Journal{\PRD}{58}{096016}{1998}; \Journal{\PRD}{60}{093001}{1999};
    J.N. Bahcall, hep-ph/0002018 (Feb, 2000);
    M.C. Gonzalez-Garcia, P.C. de Holanda, C. Pena-Garay and J.W.F. Valle,
    \Journal{\NPB}{573}{3}{2000}.
\bibitem{Homestake}
	B. T. Cleveland. et. al., \Journal{\APJ}{496}{505}{1998}. 
\bibitem{SAGE}
	SAGE Collaboration, J.N. Abdurashitov. et. al., \Journal{\PRC}{60}{055801}{1999}. 
\bibitem{GALLEX}
	GALLEX Collaboration, W. Hampel et al., \Journal{\PLB}{447}{172}{1999}. 
\bibitem{GNO}
	GNO Collaboration, M. Altmann. et. al., hep-ex/0006034 (June, 2000)
\bibitem{MassiveNeutrino}
    Z. Maki, M. Nakagawa and S. Sakata, \Journal{\PTP}{28}{870}{1962}.
    See also 
    B. Pontecorvo, \Journal{\JETPUSSR}{34}{247}{1958};
    B. Pontecorvo, \Journal{\ZETP}{53}{1717}{1967};
    V. Gribov and B. Pontecorvo, \Journal{\PLBOLD}{28B}{493}{1969}.
\bibitem{RecentK2K}
    T.Ishida, hep-ex/0008047 (Aug 2000); S. Boyd, hep-ex/0011039 (Nov., 2000).
\bibitem{QVO} 
    See for example, 
    A. de Gouve\^{a}, A. Friedland and H. Murayama, 
		\Journal{\PRD}{60}{093011}{1999}; 
    	\Journal{\PLB}{490}{125}{2000}; 
    A. Friedland, \Journal{\PRL}{85}{936}{2000}; 
    G.L. Fogli, E. Lisi, D. Montanino and A. Palazzo, \Journal{\PRD}{62}{113004}{2000}.
\bibitem{NearlyBiMaximal} 
    H. Fritzsch and Z.Z. Xing, \Journal{\PLB}{372}{265}{1996}; \Journal{\PLB}{440}{313}{1998};
    M. Fukugida, M. Tanimoto and T. Yanagida, \Journal{\PRD}{57}{4429}{1998}; 
    M. Tanimoto, \Journal{\PRD}{59}{017304}{1999}.
\bibitem{Mixing}
    D. V. Ahluwalia, \Journal{\MPL}{13}{2249}{1998};
    V. Barger, P. Pakvasa, T.J. Weiler and K. Whisnant, \Journal{\PLB}{437}{107}{1998};
    A. Baltz, A.S. Goldhaber and M. Goldhaber, \Journal{\PRL}{81}{5730}{1998};
    M. Jezabek and Y. Sumino, \Journal{\PLB}{440}{327}{1998};
    R.N. Mohapatra and S. Nussinov, \Journal{\PLB}{441}{299}{1998};
    Y. Nomura and T. Yanagida, \Journal{\PRD}{59}{017303}{1999};
    Q. Shafi and Z. Tavartkiladze, \Journal{\PLB}{451}{129}{1999};
    \Journal{\PLB}{482}{145}{2000};
    I. Starcu and D.V.Ahluwalia, \Journal{\PLB}{460}{431}{1999};
    C.H. Albright and S.M. Barr, \Journal{\PLB}{461}{218}{1999};
    R.N. Mohapatra, A. P\'{e}rez-Lorenzana and C. A. de S. Pires, \Journal{\PLB}{474}{355}{2000}.
\bibitem{SU3U1}
    F. Pisano and V. Pleitez, \Journal{\PRD}{46}{410}{1992};
    P.H. Frampton, \Journal{\PRL}{69}{2889}{1992};
    D. Ng, \Journal{\PRD}{49}{4805}{1994}.
    See also,
    M. Singer, J.W.F. Valle and J. Schechter, \Journal{\PRD}{22}{738}{1980}.
\bibitem{SU3U1Topics}
    R. Barbieri and R.N. Mohapatra, \Journal{\PLB}{218}{225}{1989};
    J. Liu, \Journal{\PLB}{225}{148}{1989};
    R. Foot, O. F. Hern\'{a}ndez, F. Pisano and V. Pleitez, \Journal{\PRD}{47}{4158}{1993};
    V. Pleitez and M.D. Tonasse, 
		\Journal{\PRD}{48}{2353}{1993}; 
    	\Journal{\PRD}{48}{5274}{1993}; 
		\Journal{\PLB}{430}{174}{1998};
    P.H. Frampton, P.I. Krastev and J.T. Liu, \Journal{\MPL}{9}{761}{1994};
    F. Pisano, V. Pleitez and M.D. Tonasse, hep-ph/9310230 v2 (Feb, 1994);
    F. -z. Chen, \Journal{\PLB}{442}{223}{1998};
    N.A. Ky, H.N. Long and D.V. Soa, \Journal{\PLB}{486}{140}{2000}.    
    See also,
    J.W.F. Valle and M. Singer, \Journal{\PRD}{28}{540}{1983}.
\bibitem{Lately}
	M.B. Tully and G.C. Joshi, hep-ph/9810282.v2 (Jan,1999), hep-ph/0011172 (Nov., 2000);
	J.C. Montero, C.A. Pires and V. Pleitez, hep-ph/0011296 (Nov., 2000) to apear in Phys. Lett. B(2001);
	hep-ph/012096 v2(Mar., 2001);
	T. Kitabayashi and M. Yasu\`{e}, in preparation. 
\bibitem{Kita00a}
    T. Kitabayashi and M. Yasu\`{e}, hep-ph/0006040 (June, 2000) to be published in Phys. Rev. D {\bf 63} (2001).
	See also, Y. Okamoto and M. Yasu${\grave {\rm e}}$, \Journal{\PLB}{466}{267}{1999}.
\bibitem{Kita00b}
    T. Kitabayashi and M. Yasu\`{e}, hep-ph/0010087 (Oct, 2000) to be published in Phys. Rev. D {\bf 63} (2001).
\bibitem{Kita01a}
    T. Kitabayashi and M. Yasu\`{e}, hep-ph/0102228 (Feb, 2001).
\bibitem{Lprime}
    R. Barbieri, L.J. Hall, D. Smith, N.J. Weiner and A. Strumia, \Journal{\JHEP}{12}{017}{1998}.
\bibitem{EarlierLprime}
    S.T. Petcov, \Journal{\PLBOLD}{110B}{245}{1982};
    C.N. Leung and S. T. Petcov, \Journal{\PLBOLD}{125B}{461}{1983};
    A. Zee, \Journal{\NPBOLD}{264B}{99}{1986}. 
\bibitem{1-loopLprime}
    C. Jarlskog, M. Matsuda, S. Skadhauge and M. Tanimoto, \Journal{\PLB}{449}{240}{1999};
    P.H. Frampton and S. L. Glashow, \Journal{\PLB}{461}{95}{1999}.
\bibitem{2-loopLprime}
    L. Lavoura, \Journal{\PRD}{62}{093011}{2000};
    T. Kitabayashi and M. Yasu\`{e}, \Journal{\PLB}{490}{236}{2000}
\bibitem{1loop2loop}
	L. Bento and J.W.F. Valle, \Journal{\PLB}{264}{373}{1991};
	K.S. Babu, R.N. Mohapatra and L. Rothstein, \Journal{\PRD}{45}{5}{1992};
    J.T. Peltoniemi, A.Yu. Smirnov and J.W.F. Valle, in Ref. \cite{2-loop};
    J.T. Peltoniemi, D. Tommasini and J.W.F. Valle, \Journal{\PLB}{298}{383}{1993}; 
    J.T. Peltoniemi and J.W.F. Valle, \Journal{\NPB}{406}{409}{1993}. 
\bibitem{1loop2loopNew}
    A.S. Joshipura and S. D. Rindani, \Journal{\PLB}{464}{239}{1999}; 
    D.Chang and A. Zee, \Journal{\PRD}{61}{071303}{2000}.
\bibitem{HeavyE}
	For example, M.\"{O}zer, \Journal{\PRD}{54}{1143}{1996}..
\bibitem{1-loop}     
    A. Zee, \Journal{\PLBOLD}{93B}{389}{1980}; \Journal{\PLBOLD}{161B}{141}{1985};
    L. Wolfenstein, \Journal{\NPB}{175}{93}{1980}.
\bibitem{2-loop}
    A. Zee, in Ref.\cite{EarlierLprime};
    K.S. Babu, \Journal{\PLB}{203}{132}{1988};
    D. Chang, W-Y. Keung and P.B. Pal, \Journal{\PRL}{61}{2420}{1988};  
    J.T. Peltoniemi, A.Yu. Smirnov and J.W.F. Valle, \Journal{\PLB}{286}{321}{1992};
    See also D. Choudhury, R. Gandhi, J.A. Gracey and B. Mukhopadhyaya, \Journal{\PRD}{50}{3468}{1994}. 
\bibitem{FCNCSU3}
    J.C.Montero, F.Pisano and V.Pleitez, \Journal{\PRD}{47}{2918}{1993};
    M.\"{O}zer, \Journal{\PRD}{54}{4561}{1996}.
\bibitem{FCNC}
    S.L.Glashow and S.Weinberg, \Journal{\PRD}{15}{1958}{1977};
    H.Georgi and A.Pais, \Journal{\PRD}{19}{2746}{1979}.
\end{references}
\end{document}